\documentclass{jps-cp}
\usepackage{txfonts} 
\usepackage{verbatim}
\title{$x_{\textrm{opt}}$ and $x_{\textrm{c}}$  in Cuprate 
Superconductors}

\author{J{\"u}rgen \textsc{R{\"o}hler}}

\inst{Fachgruppe Physik, Universit{\"at} zu K{\"oln}, 50937 K{\"oln}, Germany}

\email{juergen.roehler@uni-koeln.de}

\recdate{September 16, 2019}%

\abst{The characteristic hole concentrations $x_{\textrm{opt}}\simeq 0.16$ and $x_{\textrm{c}}\simeq 0.19$ in the under- to overdoped transition regime of cuprate superconductors are shown to be intimately interelated by mesoscale organization of bond-like RVB arrays extending over $3\times 4$ copper sites and comprising a hole pair spaced by $3a_{0}$. The columnar organization maximizes the pairing gap but reduces the superconducting stiffness, while the zigzag organization maximizes the superconducting stiffness at reduced $T_{\rm c}$. The latter goes hand in hand with the closure of the pseudogap. Columnar and zigzag mesoscale organizations of the bond-like RVB arrays are displacively transformed in each other between $x_{\textrm{opt}}$ and $x_{\textrm{c}}$.

}

\kword{superconductivity, cuprates, RVB, pairing, pseudogap}

\begin{document}
\maketitle

\section{Introduction}
Two distinct  
hole concentrations, $x_{\rm opt}\simeq 0.16$ and $x_{\rm c}\sim 0.19$, 
characterize the transition from the under- to the overdoped regime 
of cuprate superconductors. 
At $x_{\rm opt}$ the critical temperature $T_{\rm c}(x)$ reaches its maximum value.  
Fundamental properties of the superconducting phase, such as the zero-temperature superfluid density, the condensation energy, the zero temperature critical Silsbee field, however, are maximal only at 
$x_{\rm c}$ where the critical temperature is 
10\% below its maximum value (``slightly overdoped''). $x_{\rm c}$ marks also the closure of a pseudogap which arises in the normal phase at elevated temperatures right up to the thermal  decomposition of the doped crystals. Thermodynamic and magnetic data indicate that the high temperature density of states near the chemical potential $\mu = E_{\rm F}$ is partially depleted \cite{lor2001}. Lowering $T$ the number of 
available states is significantly reduced at $T^*(x) > T_{0}(x) > 
T_{\rm c}(x)$, a characteristic temperature used to define the energy scale of the pseudogap, corroborated by the uniform magnetic susceptibility $\chi(T, x)$, 
the spin susceptibility $\chi_{\rm s}(T, x)$ (NMR Knight shift) and many   
spectroscopic data.   The onset temperature 
of nonlinear diamagnetic susceptibility from superconducting phase 
fluctuations is marked by $T_{0}(x)$, and the dome-shaped onset of superconducting amplitude 
fluctuations by $T_{\rm c}(x)$.

$T^*(x)$ decreases linearily with increasing doping 
and falls to zero at $x_{\rm c}$.  The pseudogap scale $\Delta/k_{B} = J(1-x/x_{c})$ with $J\simeq 1000$~K  arises from both, spin and charge degrees of freedom, and is 
comparable to the antiferromagnetic superexchange energy.    
Notably this strongly entropy-depleting pseudogap is completely absent 
beyond $x_{\rm c}$ throughout the strongly overdoped regime up to $x\simeq 0.25$ 
where superconductivity vanishes.   

The apparent disparity of the superconducting properties at  $x_{\rm opt}$ and $x_{\rm c}$ still stirs lively debates, particularily on a possible quantum critical point at $x_{\rm c}$. In this short note 
we attempt to outline, almost phenomenologically in the framework of the resonating valence bond (RVB) theory\cite{and2011}, that both characteristic hole concentrations are intimately interrelated through the spatial structure of a non-local $d$-wave RVB pair state constrained by strong off-site hole -- hole repulsion.  We show that geometrically limited number of connections among such $d$-wave RVB pair states is most likely at the origin of the pseudogap. Weak overdoping of columnar closest packed but partially unconnected RVB pair states is demonstrated to induce a displacive transition into a  fully connected zigzag structure.

%

\section{{\it d}-wave RVB states}

\subsection{Electron-hole asymmetry}
The physical properties of the hole doped cuprate superconductors 
are in essence governed by relaxation of the frustrated 
kinetic energy causing antiferromagnetic ``kinetic'' exchange  
in an insulating parent compound where spin-half electrons are 
kept apart by large on-site (Mott - Hubbard) Coulomb repulsion 
$U_{d}$\cite{and1997}. The electronic excitations and the pairing interaction 
in such a liquid of short range spin singlet electron pairs
have been elaborated with Anderson's theory of resonating 
valence bonds (RVB)\cite{and2011}, mostly under the implicit assumption of electronic homogeneity.

The strong desire of lattice electrons 
to delocalize themselves favores homogeneous spatial distributions. 
In general metals prefer most densely packed lattice structures optimizing the gain of (negative) kinetic energy and opposing segregational interactions. The normal electron liquid in doped cuprate superconductors however appears to be intrinsically strained, even in most carefully annealed stoichiometric compounds with relatively simple crystal lattices. Notorious chemical metastability and electronic symmetry breaking seem to be constituitive properties of superconducting cuprates.  One  important reason is the intrinsic electron -- hole asymmetry\cite{and2011} arising from the on-site electron -- electron repulsion $U_{d}$, another off-site hole -- hole repulsion rendering electron -- electron and hole -- hole pairs spatially asymmetric. 

In the narrow doping range between $x_{\textrm{opt}}$ and $x_{\textrm{c}}$ however 
the strain in the electron liquid seems to be gone, with them also doping induced lattice 
effects\cite{kal1997, boz2000}.  Here the superconducting RVB {\it d}-wave function 
is found to get entirely stiff.

\subsection{Off-site hole -- hole repulsion}
Holes doped into a RVB liquid of short range spin singlets are empty states
assumed to move free.  Though, double occupancy of hole sites is
excluded by $U_{d}$ as it is for electrons at electron sites.  It is notable that the 
mobile empty states in the CuO$_{2}$ lattice are created by singlet
combination of the symmetrized orbital on the four oxygens around a
Cu$^{2+}$.  Such holes are virtual Cu$^{3+}$
sites with a hard core diameter larger 
than the lattice constant $a_{0}$, set by the ``cage" of the oxygen environment. 
The energy of this so-called Zhang-Rice (ZR) state is by $\sim
5$~eV distinctly lower than that of all other combined states, thus
such mobile hole states are stable throughout the entire relevant doping
regime, $0< x\leq 0.33$.  
Coulomb repulsion between oxygen $2p_{x,y}$ states at the connecting
site of two ZR singlets, $U_{p}\simeq 6$~eV, may be naively
held responsible for the supression of nearest neighbor hole pairs.  A
rough estimate for two ZR states built out of Wannier states
yields a repulsion $R\simeq U_{p}/32 \simeq 0.19$~eV\cite{bar1994}.
More detailed theoretical work on this issue is elusive.

\begin{figure}[tbh]
\includegraphics[width=15.5cm]{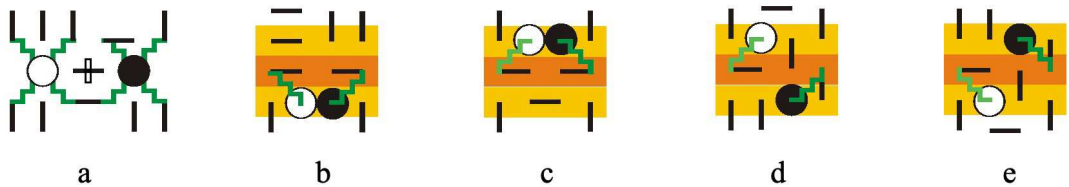}
\caption{(Color online) Adiabatic snapshots of a resonant valence bond configuration showing singlet pairs of electrons spaced by one lattice constant $a_{0}$ (horizontal and vertical black bars), and two holes doped at electron sites with opposite spins (large white and black circles). 
The holes spaced by $3a_{0 }$  along the Cu -- O bond axis experience directional attraction (orange bar) through kinetic exchange with the electron singlets in an array of as many as $3\times 4$ copper sites (yellow background). Off-site hole -- hole repulsion excludes occupancy of nearest neighbor sites (see text). Hence the odd $a_{0}$ hole -- hole pairing symmetry realizes only for spacings 
$\geq 3a_{0 }$ with the location of the inversion center for all directional space and spin coordinates at the central intervening oxygen site (small rectangle in the center of {\bf a}). Green wavy lines indicate spin conserving next-nearest neighbor electron -- hole interactions ($t'$) along the lattice diagonals (nodal lines in $k$-space).  {\bf b--e} display two - hole excitations of {\bf a}. {\bf d} and {\bf e}: The diagonal excitations in sites from antiparallel electrons avoid off-site repulsion but do not lock spin and charge. 
{\bf b} and {\bf c}: Virtual excitations to excluded nearest neighbor hole -- hole sites from antiparallel electrons locking spin and charge. 


}
\label{f1}
\end{figure}

Hole doping contracts strongly the basal plane area. A preliminary inspection of the 
nonlinearities in the doping dependence of basal lattice constants, bulging at $x_{\rm opt}$ 
and collapsing at $x_{\rm c}$, lead us to suggest that the hard cores of the virtual Cu$^{3+}$
holes exclude not only onsite double occupancy but also nearest neighbor hole -- hole configurations\cite{roe2004}. 
Hence empty replica of electron -- electron spin singlets, spaced by $a_{0}$, have to be 
excluded from the possible hole -- hole configuration manifolds of the doped electron system.  
Therefore also clusters of nearest neighbor holes, hole stripes, hole puddles {\it etc.}

Indirect experimental support for this electron -- hole asymmetry, 
particularily by STM\cite{koh2012}, comes from the charge density modulations (CDM) 
evidenced  at pseudogap excitation energies in the underdoped regime.  
These CDM's are sinusodial in one spatial dimension (bidirectional in $x, y$) with a commensurate period 
$4a_{0}$ centered at oxygen sites. 
We view the stick-like CDM patterns (see inset in Fig.\ref{f2}) as signatures of bond-centered hole -- hole exchange over $3a_{0}$, extending effectively to $4a_{0}$ due to the large hole sizes.

\subsection{Bond-like RVB arrays}

The frustrated kinetic energy causing antiferromagnetism in the 
parent compound is relaxed through interactions which are short range in 
space and time. Fig.\ref{f1}a displays schematically the locality of an isolated doped {\it d}-wave RVB state at pseudogap excitation energies. Bond-centered hole -- hole pairing in the liquid of electron singlets arises for spacings of $3a_{0}$ along the Cu -- O bonds (antinodal {\it loci} in $k$-space) from an axial  inversion center for all directional space and spin coordinates at the central intervening oxygen site (small rectangle in the center of Fig.\ref{f1}a). Figs.\ref{f1}b--e illustrate the kinetic exchange interaction effectively attracting the holes.  Hole -- hole configurations spaced by even multiples of the lattice constant, and site centered holes spaced by $a_{0}\sqrt{5}$ are unable to lock the spin and charge degrees of freedom in a pairing state. 

The green wavy lines in Fig.\ref{f1} symbolize quantum fluctuations along the lattice diagonal 
exchanging holes with electrons at the next-nearest neighbor sites without spin flips. The remote hybridization of the Cu$4s$ with the Cu$3d$O$2p$ orbitals promotes nodal fluctuations such that the ratio of the spin conserving, and the spin flipping hopping matrix element $-t'/t$ achieves crucial importance for the pairing mechanism.

\begin{figure}[tbh]
\includegraphics[width=15.5cm]{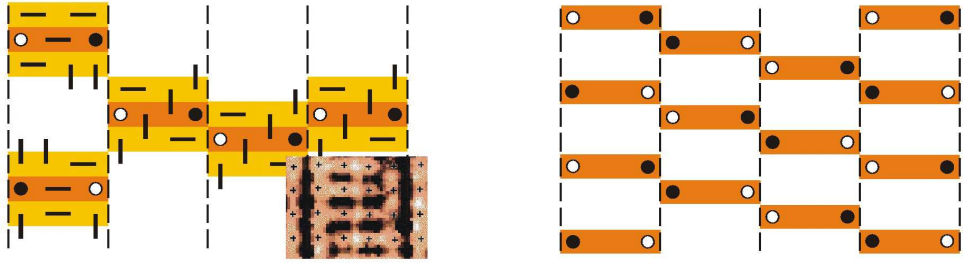}
\caption{(Color online) Adiabatic snapshots of columnar connected bond-like RVB arrays. Orange bars indicate directional hole -- hole attraction in the yellow colored array of RVB electron singlet states. {\it Left:} Underdoped occupation. The inserted pattern are a time integated record taken by STM at pseudogap energies\cite{koh2012}.  {\it Right:} Optimally doped  occupation (electron singlet states omitted)

}
\label{f2}
\end{figure}

We expect the nodal fluctuations to generate effectively axial hole -- hole  
attraction by kinetic exchange in an RVB array with at least 
$3\times 4=12$ copper sites, see Fig.\ref{f1} and caption. The size of the resulting pairing gap $\Delta_{\rm p}$ will be almost independent on the average hole concentration, and larger than the energy of the superconducting amplitude fluctuations $k_{\rm B}T_{c}$.  
The spatial extension of  the bond-like RVB array sets the full filling of the CuO$_2$ lattice with paired holes to  $x=1/6\simeq 0.16=x_{\rm opt}$, see Fig.\ref{f2} ({\it right}). Hence ``overdoping'' may only accomodate additional holes (still constrained by off-site repulsion) in ferromagnetic site-centered hole -- hole configurations, which will destroy the antiferromagnetic pairing symmetry of the kinetic exchange, see Fig.\ref{f4} ({\it right}). Thus three holes, still constrained by off-site repulsion even in the overdoped regime, will depair a bond-like RVB array. Complete destruction of the pairing gap $\Delta_{\rm p}$ will occur at $x=3/12 = 0.25$. The unusual electronic properties of the overdoped regime, {\it e.g} the paramagnetic susceptibility, may have its origin in a two-component fluid of paired bond-like RVB arrays and free electrons. It is then reasonable to expect that the superconducting stiffness decreases linearily also in the overdoped regime\cite{mah2019}. 

\subsubsection{Nano-sized columns}

These bond-like RVB arrays  break the rotational symmetry of 
the lattice and give rise to uniaxial (``nematic") strain in the electronic lattice spreading macroscopically 
with increasing hole concentrations. Well known macroscopic deformable bodies favor 
relaxation of uniaxial microscopic strain  by formation of meso-scale unidirectional areas, and orientating them most appropriately. Large field STM maps\cite{koh2012} of underdoped samples evidence nanometer-scale unidirectional clusters in  $x$ and $y$ direction with no preference of one over the other. 
Apparently doping in the $C_{4}$ symmetric  RVB liquid proceeds as 
correlation of doped holes in nano-sized clusters with columns of  bond-like RVB arrays. 
The interconnection of bond-like RVB arrays of neighbored columns underlies the 
off-site occupation constraint as each single bond-like RVB array. Fig.\ref{f2} ({\it left})
displays schematically an adiabatic snapshot of an underdoped, {\it i.e.} only partially occupied columnar cluster of  bond-like RVB arrays. The STM pattern exhibits clearly the bond-like structure of interchanging holes and the intercolumnar repulsion with periodicity $4a_0$.  
Naturally it does not yield information on the spins and their dynamics and not on the dynamics of the hole--hole exchange; its integrated intensity may be taken as a measure of the average number of holes\cite{fuj2014}.

\begin{figure}[tbh]
\includegraphics[width=15.5cm]{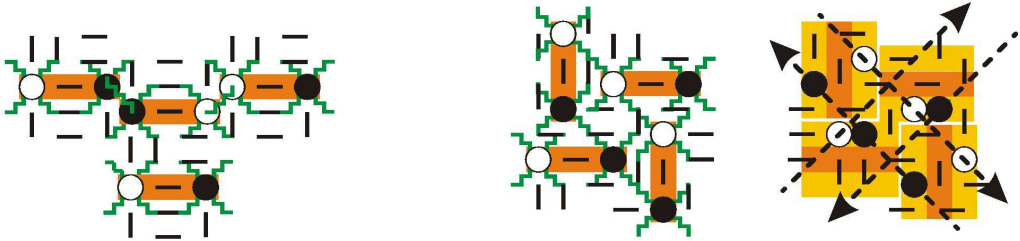}
\caption{(Color online) {\it Left}: Nodal connections between bond-like RVB arrays in columnar  organization. {\it Middle}: Nodal connections between bond-like RVB arrays in zigzag organization. 
{\it Right}: Zigzag organization; collective hole displacements along the lattice diagonal generate the excited configurations of bond-like RVB arrays displayed in Fig.\ref{f1} b -- e.

}
\label{f3}
\end{figure}

\subsubsection{Zigzag packing}
The organization of the bond-like RVB arrays on mesoscopic and macroscopic scales encounters  a  connectivity problem.  Fig.\ref{f3} schematizes the nodal electron -- hole quantum fluctuations (green wavy lines) in different most closely packed configurations: columnar ({\it left}), and `tweedy" zigzag ({\it middle}) patterns. Inter-columnar electron -- hole fluctuations connecting nearest neighbored bond-like RVB arrays are obviously geometrically constrained by "ferro" equal spin hole -- hole paths. Intra-columnar  electron -- hole fluctuations however are able to connect via common electron spin singlet "antiferro" paths.

\begin{figure}[tbh]
\includegraphics[width=15.5cm]{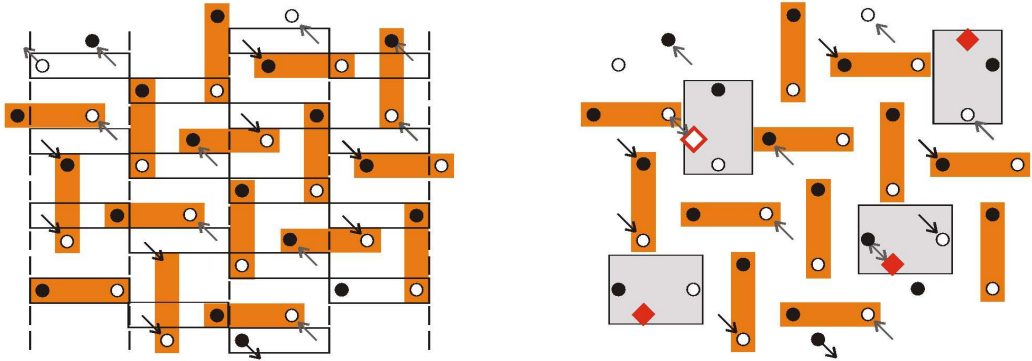}
\caption{(Color online) {\it Left}: Displacive transformation of optimally doped and columnar organized  
bond-like RVB arrays (white bars) into the optimally connected zigzag organization. The arrows mark the directions of the collective hole displacements. 
{\it Right}: Weakly overdoped zigzag organization of bond-like RVB arrays at $x\simeq 0.19$. The overdoped excess holes (red diamonds) depair the bond-like RVB arrays but do not destroy the zigzag organization of the remaining. Note possible ordering of the depaired sites.

}
\label{f4}
\end{figure}

The fully packed columnar configuration may be displacively transformed into the energetically more favorable zigzag configuration, see Fig.\ref{f4} ({\it left}). For that columnar organized holes have to be collectively displaced along the lattice diagonals to next nearest neighbor sites, while reversing the displacive shifts from striation to striation. Note the arrows from the hole sites in the columnar configuration (open rectangles). This transformation of  a ``martensitic" type creates a most densely packed zigzag configuration in which all bond-like RVB arrays are able to connect through common electron spin singlet states. Clearly this is the energetically superior configuration of the two enhancing electron delocalization through unconstrained long range phase correlation. Nevertheless the columnar configuration may still survive at $x_{\rm opt}$ with most densely packed bond-like RVB states, because uniaxial microscopic strain inhibits long range collective displacements. Thus the number of pair states is maximal, but the superconducting stiffness stays below its optimal value. 

%

\section{Closure of the pseudogap at $x_{\rm c} > x_{\rm opt}$}

The columnar organization of  $x_{\rm opt}$ at $x_{\rm opt}$  is apparently metastable. A tiny amount of overdoping excess holes however may transform the nano-sized colums into the fully connected zigzag configuration, however with the consequence of partial pair breaking by "ferro" excess holes in some bond-like RVB arrays.  This situation is illustrated in Fig.\ref{f4} ({\it right}). The "ferro" holes are marked by open and closed diamonds for spin up and spin down electron site, and local depairing by the grey rectangles around the broken hole pairs (omitted orange bar). Certainly local depairing will affect the superconducting amplitude fluctuations decreasing $T_{\rm c}$, however at the same time establishes unconstrained long range phase coherence thus maximal superconducting stiffness. Hence superconductivity at $x_{\rm c}$ turns out to be more robust than at $x_{\rm opt}$, though at a diminuished value of $T_{\rm c}$.

Hence $x_{\rm opt}$ at $x_{\rm opt}$ are intimately connected with each other through the geometrical constraints governing the spatial organization of fully coherent assembly of nano-sized bond-like RVB arrays.

\end{document}